
\magnification 1200
\baselineskip=18pt
\null\vskip -0.5truein
\hsize=5.5truein
\vsize=9.0truein
\overfullrule=0pt
\def\r{\vbox{\hbox{\raise.90mm\hbox{$>$}}
\kern-18pt\hbox{\lower.90mm\hbox{$\sim$}}}}
\def\l{\vbox{\hbox{\raise.90mm\hbox{$<$}}
\kern-18pt\hbox{\lower.90mm\hbox{$\sim$}}}}

\line{\hfill{CERN-TH.7362/94}}
\line{\hfill{CTP-TAMU-37/94}}
\line{\hfill{NUB-TH-3098/94}}
\centerline{\bf NEUTRALINO EVENT RATES IN DARK MATTER DETECTORS}
\bigskip
\centerline{R. Arnowitt}
\bigskip
\centerline{Center for Theoretical Physics, Department of Physics}
\centerline{Texas A\&M University, College Station, TX  77843-4242}
\centerline{Pran Nath}
\centerline{Theoretical Physics Division, CERN, CH-1211 Geneva 23}
\centerline{and}
\centerline{\footnote*{Permanent address}Department of Physics,
Northeastern University, Boston, MA  02115}
\bigskip

\centerline{ABSTRACT}
\smallskip

The expected event rates for ${\tilde Z_{1}}$ dark matter for a variety of dark
matter detectors are studied over the full parameter space with tan $\beta\leq$
20 for supergravity grand unified models. Radiative breaking constraints are
implemented and effects of the heavy netural Higgs included as well as loop
corrections to the neutral Higgs sector.  The parameter space is restricted so
that the ${\tilde Z_{1}}$ relic density obeys 0.10 $\leq\Omega_{\tilde
Z_{1}}h^2\leq 0.35$, consistent with the COBE data and astronomical
determinations of the Hubble constant.  It is found that the best detectors
sensitive to coherrent ${\tilde Z_{1}}$ scattering (e.g. Pb) is about 5-10 more
sensitive than those based on incoherrent spin dependent scattering (e.g. CaF).
In general, the dark matter detectors are most sensistive to the large tan
$\beta$ and small $m_o$ and $m_{\tilde g}$ sector of the parameter space.
\vfill\eject

\noindent
1.  INTRODUCTION
\smallskip

There is much astronomical evidence that more than 90\% of our Galaxy, and
perhaps of the universe is made up of dark matter of unknown type.  In
galaxies, this matter has been detected by its gravitational effects on the
motion of stars and gas clouds.  A large number of candidates for dark matter
have been suggested both from astronomy and particle physics.  In this paper we
will limit our discussion to supersymmetry models with R parity, as they offer
a natural candidate for dark matter, the lightest supersymmetric particle (LSP)
which is absolutely stable.  Thus the relic LSP left over from the big bang
could be the dark matter present today.  Further, in supergravity GUT models,
for almost all the parameter space of most models, the LSP is the lightest
neutralino, the ${\tilde Z_{1}}$.  (The alternate possibility, that the
sneutrino is lightest occurs only rarely.)  Thus we will consider here only the
${\tilde Z_{1}}$ dark matter candidate, and do so within the framework of
supergravity grand unification with radiative breaking.

In this paper we discuss the expected event rates for a number of dark matter
detectors using the following nuclei:  $^3He$, $^{40}Ca~^{19}F_2$, $^{76}Ge
+^{73}Ge$, $^{79}Ga~^{75}As$, $^{23}Na^{127}I$ and $^{207}Pb$. The first two
represent nuclei which are most sensitive to spin dependent incoherrent
scattering of ${\tilde Z_{1}}$ by the nuclei, while the last four are
increasingly sensitive to coherrent scattering.  Pb could be a candidate for a
superconducting detector.

A great deal of work has already been done on the question of dark matter
detector rates [1-7].  We present here an analysis over the entire SUSY
parameter space with tan $\beta\leq$20 which takes into account several
important effects not generally treated before:

\item{$\bullet$} Radiative breaking.  Almost all previous analysis has been
done within the framework of the MSSM which does not include the constraints of
radiative breaking of SU(2)xU(1).  These constraints allow the determination of
$\mu^2$ and $m_A$ ($\mu$ is the $H_1-H_2$ Higgs mixing parameter, A is the CP
odd Higgs boson) in terms of the other parameters.  (Some previous analyses
have varied $m_A$ arbitarily, obtaining spuriously large event rates.)

\item{$\bullet$} As pointed out in Refs. [6,7] the heavy neutral Higgs, H, can
make an important contribution to the event rates.  We have included this for
the entire parameter space and find that the H contribution relative to the
light Higgs, h, can range from 1/10 to 10 times as large.

\item{$\bullet$} As is well known, loop corrections to $m_h$ are important due
to the fact that the t quark is heavy [8].  We have also included the loop
correction to ${\tilde\alpha}$ (the rotation angle arising in diagonalizing the
h-H mass matrix).  These actually cancel much of the effects of the loop
corrections to $m_h$.

\item{$\bullet$} The COBE constraints on the ${\tilde Z_{1}}$ relic density are
included.  This strongly limits the region of SUSY parameter space that is
allowed.  In calculating these relic density constraints it is essential to
include the effects of the h and Z s-channel poles [9-11] for gluinos with mass
$m_{\tilde g}\l$ 450 GeV.

There are several effects we have not included here.  Most noteworthy are that
we have ommitted the possible WW, ZZ, Zh, hh final states in the ${\tilde
Z_{1}}$ annihilation for the relic density calculation (which can occur when
$m_{\tilde Z_{1}}$ gets to the upper end of its allowed spectrum i.e.
$m_{\tilde Z_{1}}\r M_W$ and we have followed Refs. [12,13] in calculating the
relic density.  We estimate that this may lead to a (25-30)\% error in the
relic density, and since we have been reasonably generous in the allowed values
for the relic density, we expect this will not significantly change our final
conclusions.  We also discuss below the sensitivity of the results to changes
in the allowed region of ${\tilde Z_{1}}$ relic density.
\medskip
\noindent
II.  RELIC DENSITY CONSTRAINT
\smallskip
The COBE data suggests that dark matter is a mix of cold dark matter, CDM,
(which we are assuming here to be the relic ${\tilde Z_{1}}$) and hot dark
matter, HDM (possibly massive neutrinos) in the ratio of 2:1.  In addition
there may also be baryonic dark matter, B, (possibly brown dwarfs) of amount
$\l$ 10\% of the total.  Defining $\Omega_i = \rho_i/\rho_c$, where $\rho_i$ is
the mass density of the i$^{th}$ constituent and $\rho_c =3H^2/(8\pi G_N)$ [H =
Hubble constant, $G_N$ = Newtonian constant] is the critical mass density to
close the universe, then the inflationary scenario requires $\Sigma\Omega_i=1$.
A reasonable mix of matter is then $\Omega_{\tilde Z_{1}}\simeq 0.6$,
$\Omega_{HDM}\simeq 0.3$ and $\Omega_B\simeq 0.1$.  What can be calculated
theoretically is $\Omega_{\tilde Z_{1}} h^2$ where h = H/(100 km/s Mpc).
Astronomical observations give h = 0.5-0.75.  Thus we are lead to the estimate
\smallskip
$$\Omega_{\tilde Z_{1}}h^2\cong 0.10-0.35\eqno(1)$$

\smallskip
Eq.(1) strongly resticts the allowed SUSY parameter space, and thus it is
necessary to have a satisfactory method of calculating $\Omega_{\tilde
Z{_1}}h^2$.  (We will discuss below the effects of varying the maximum and
minimum values of $\Omega_{\tilde Z_{1}}h^2$.)  To do this, we use supergravity
GUT models [14].  These models have the advantage of being consistent with the
LEP results on unification of couplings at $M_G\simeq 10^{16}GeV$ [15], and
generate naturally spontaneous breaking of supersymmetry in a hidden sector.
In addition, by use of the renormalization group equations (RGE), the
supersymmetry breaking interactions at $M_G$ produce naturally spontaneous
breaking of SU(2)xU(1) at the elctroweak scale $M_Z$.  In general, the low
energy supersymmetry theory depends on only four parameters, $m_o$, $m_{\tilde
g}$, $A_t$, tan$\beta$, and the sign of $\mu$.  Here $m_o$ is the universal
mass of all scalar fields at $M_G$, $A_t$ is the t-quark cubic soft breaking
parameter at the electroweak scale, and tan $\beta = \langle H_2\rangle/\langle
H_1\rangle$ where $\langle H_{2,1}\rangle$ gives masses to the (up, down)
quarks.

The above may be contrasted with the MSSM (the formalism most dark matter
calculations use) which possesses no theoretical mechanism for SUSY or
SU(2)xU(1) breaking and is generally parameterized by 20 aribtrary constants.
In the supergravity models, all properties of the 32 SUSY particles (masses,
widths, cross sections, etc.) are determined in terms of the four basic
parameters and one sign.  In particular, this means that $m_A$ and $\mu$ are so
determined and are not free parameters (as usually assumed in the MSSM).
Further, one finds throughout most of the parameter space the following
(approximate) relations [16]:

$$2m_{\tilde Z_{1}}\cong m_{\tilde Z_{2}}\cong m_{\tilde W_{1}}\simeq({1\over
4}- {1\over 3}) m_{\tilde g},\eqno(2)$$
\smallskip
\noindent
while $m_h\l 130 GeV$, $m_H^2>>m_h^2$ and tan $\beta > 1$.  (Here, ${\tilde
W_{1,2}}$ are the two charginos and ${\tilde Z_{1,2,3,4}}$ are the four
neutralinos).  These relations will be important in understanding the results
below.

The calculation of $\Omega_{\tilde Z_{1}}h^2$ now proceeds in a standard
manner. Using the RGE, we first express all SUSY masses and couplings in terms
of the four basic parameters.  This is done for the parameter space over the
range

$$150 GeV\leq m_{\tilde g}\leq 1 TeV; 100 GeV \leq m_o \leq 1 TeV; -2\leq
A_t/m_o\leq 6; 2\leq tan \beta\leq 20\eqno(3)$$

\smallskip
\noindent
with a mesh $\Delta m_o$ = 100 GeV, $\Delta m_{\tilde g}$ = 25 GeV,$\Delta
(A_t/m_o)$ = 0.5, and $\Delta (tan \beta)$ = 2 or 4.  We assume a top quark
mass of $m_t$ = 167 GeV, and LEP and CDF bounds are imposed on the SUSY
spectrum. The $A_t$ range stated above exhauts the parameter space.  Note that
our analysis does not assume any specific grand unification group but only that
it is $\alpha_1\equiv (5/3)\alpha_Y$ that unifies at $M_G$.  in the early
universe, the ${\tilde Z_{1}}$ is in equilibrium with quarks, leptons, etc.
When the annihilation rate falls below the expansion rate, ``freezeout" occurs
at temperature $T_f$.  The ${\tilde Z_{1}}$ then continues to annihilate via
s-channel h and Z poles (${\tilde Z_{1}} + {\tilde Z_{1}}\rightarrow h$,
$Z\rightarrow q{\bar q}$; $\ell\bar\ell$; etc.) and t and u-channel squark and
slepton poles.  The relic density at present time is given by [13]:

$$\Omega_{\tilde Z_{1}}h^2\cong 2.4\times 10^{-11}\left ({T_{\tilde Z_{1}}\over
T_\gamma}\right )^3\left ({T\gamma\over 2.73}\right )^3 {N_f\over
J{(x_f)}}\eqno(4)$$ \smallskip
\noindent
where $N_f$ is the effective number of degrees of freedom, ($T_{\tilde
Z_{1}}/T_\gamma)^3$ is the reheating factor and

$$J(x_f) = \int ^{x_f}_o dx < \sigma v>;~~ x = kT/m_{\tilde Z{_1}}\eqno(5)$$
\smallskip
Here $\sigma$ is the annihilation cross section, {\it v} is the relative
velocity and $< >$ means thermal average.  Since annihilation occurs
non-relativistically, $x_f\approx 1/20$, one may take the thermal average over
a Boltzman distribution.  However, as stressed in Refs. [9,10,11] one cannot
generally make the non-relativisitic expansion $\sigma v = a+bv^2+...$ due to
the presence of the narrow h and Z s-channel poles.  Thus calling
$\Omega_{approx}$ the evaluation using the low {\it v} expansion, and $\Omega$
the rigorous result, we find for $\mu > 0$ that the relation 0.75 $\leq
\Omega_{approx}/\Omega\leq 1.25$ is satisfied for only 35 \% of the mesh points
for $m_{\tilde g} <$ 450 GeV, but for almost 100 \% for $m_{\tilde g}>$ 450
GeV.  The reason for this can be seen from Eq. (2).  One is close to an
s-channel pole when 2 $m_{\tilde Z{_1}} \approx {1\over 3} m_{\tilde g}$ is
near $m_h$ or $M_Z$.  Since $m_h\l 130$ GeV, this cannot happen when $m_{\tilde
g} \r 450$ GeV but one is usually somewhat near either an h or Z pole when
$m_{\tilde g}\l$ 450 GeV.  Thus a rigorous calculation is necessary for lower
mass gluinos.

The annihilation cross section $\sigma$ can be expressed in terms of the four
basic parameters $m_o$, $m_{\tilde g}$, $A_t$ and tan $\beta$.  Using then Eq.
(4) the region in parameter space obeying the COBE constraint of Eq. (1) can be
determined.
\medskip
\noindent
III.  EVENT RATE CALCULATION
\smallskip
Dark matter detectors see the incident ${\tilde Z_{1}}$ from effects of its
scattering on quarks in the nuclei of the detector.  This scattering proceeds
through s-channel squark poles (${\tilde Z_{1}} + q\rightarrow {\tilde
q}\rightarrow {\tilde Z_{1}}+ q$) and t-channel h, H and Z poles.  These are
some of the crossed diagrams to the annihilation diagrams appearing in the
relic density analysis.  Thus to a rough approximation, one may expect the
event rate to be large when the annihilation cross section is large i.e. when
$\Omega_{\tilde Z_{1}} h^2$ is small.  This makes results somewhat sensitive to
where the lower bound on $\Omega_{\tilde Z_{1}}h^2$ is set, and we will discuss
this below.

The scattering diagrams have been analysed by a number of people [1-7], and we
follow the analysis of Ref. [5].\footnote*{We include an extra factor of 4 in
the cross section, due to the Majorana nature of the ${\tilde Z_{1}}$, in
agreement with Ref. [7].}  One may represent the diagrams by the effective
Lagrangian

$${\cal L}_{eff} = ({\bar\chi}_1 \gamma^{\mu}\gamma^5 \chi_1){\bar
q}\gamma^{\mu}
(A_qP_L+B_qP_R)q+({\bar\chi}_1\chi_1)C_qm_q q{\bar q}\eqno(6)$$
\smallskip
\noindent
where q(x) is the quark field, $m_q$ is its mass, and $\chi_1(x)$ is the
${\tilde Z_{1}}$ field.  $A_q$ and $B_q$ arise from the Z t-channel pole and
${\tilde q}$ s-channel pole, and $C_q$ from the h, H t-channel poles and
${\tilde q}$ s-channel pole.  Expressions are given for A,B,C in Ref. [5].  The
first term of Eq. (6) give rise to spin dependent incoherrent scattering while
the second term gives rise to coherrent scattering.  There are several points
to be made concerning the latter amplitude.  In general, the ${\tilde Z_{1}}$
is a linear combination of two gauginos and two Higgsinos:

$$\chi_1 = \alpha {\tilde W_{3}} + \beta {\tilde B} + \gamma {\tilde H_{2}} +
\delta{\tilde H_{1}}\eqno(7)$$
\smallskip
\noindent
The $\alpha,\beta,\gamma,\delta$ can all be calculated in terms of the four
basic parameters, and throughout the allowed part of the parameter space one
finds

$$\beta >\alpha,\delta >>\gamma\eqno(8)$$
\smallskip
\noindent
The coefficient $C_q$ for the h and H poles is [17]:

$$C_q = {g^2_2\over 4M_W}
        \left[ \left\{
               { { { cos\tilde\alpha \over sin\beta} {F_h\over m_h^2} }
        \atop     -{ sin\tilde\alpha \over cos\beta} {F_h\over m_h^2} }
              \right\}
               \left\{
               { { { sin\tilde\alpha \over sin\beta} {F_H\over m_H^2} }
        \atop      { cos\tilde\alpha \over cos\beta} {F_H\over m_H^2} }
              \right\}
        \right] {u-quark \atop d-quark}\eqno(9)$$ \smallskip
\noindent
where $F_h = (\alpha-\beta  tan\theta_W)(\gamma cos\alpha + \delta sin\alpha)$
and $F_H=(\alpha-\beta tan\theta_W)(\gamma sin\alpha-\delta cos\alpha)$.  The
tree value of ${\tilde\alpha}$ (the rotation angle that diagonalizes the h-H
mass matrix) can be expressed in terms of the tree value of $m_h$[3].  Since
loop corrections are large for the h particle, we have also included the loop
corrections to ${\tilde\alpha}$ [8] in our calculation of $C_q$. Remarkably
though, $\tilde\alpha_{loop}[(m_h)_{loop}]$ is generally quite close (within a
few percent) to $\tilde\alpha_{tree}[(m_h)_{tree}]$.  Thus ${\tilde\alpha}$
remains small i.e. ${\tilde\alpha} = O(10^{-1}rad)$.  [Note, however, had one
just inserted the loop correction to $m_h$ into the tree formula for
${\tilde\alpha}$, one would have incorrectly obtained a large value for
${\tilde\alpha}$, i.e. ${\tilde\alpha}$ = O(1 rad.)!]  One can now see why the
H Higgs can make a significant contribution to $C_q$ even though
$m_H^2>>m_h^2$.  For the d-quarks, the h term is reduced by a factor tan
${\tilde\alpha}$ relative to the H term.  Further, the second fact in $F_h$ is
small, either because $\gamma$ is small or sin ${\tilde\alpha}$ is small.  Thus
for d-quarks, the H contribution can range from 1/10 to 10 times the h
contribution, depending on the point in the parameter space.  [For u-quarks,
the H term is generally small.]

The total event rate is given by [5]

$$R= (R_{coh + R_{inc}}) [\rho_{\tilde Z_{1}}/(0.3 GeV cm^{-3})][<v_{\tilde
Z_{1}}>/(320 km/s)] [events/kg da]\eqno(10)$$
\smallskip
\noindent
where the coherrent and incoherrent rates are

$$R_{coh} = 16 {{m_{\tilde Z_{1}} M_N^2M_Z^4}\over (M_N + m_{\tilde Z_{1}})^2}
210\zeta_{ch}\mid M_{coh}\mid^2$$

$$R_{inc} = 16 {{m_{\tilde Z_{1}}M_N\over (M_N+ m_{\tilde Z_{1}})^2}
580 \lambda^2J(J+1) \zeta (r_{sp})\mid M_{inc} \mid ^2}\eqno(11)$$
\smallskip
\noindent
Here $M_N$ is the nuclear mass, $\zeta(r_{ch}), \zeta(r_{sp})$ are charge and
spin form factor corrections, J is the nuclear spin and $\lambda$ is defined by
$<N\mid\Sigma {\buildrel{\rightarrow}\over{S_i}}\mid N >= \lambda <
N\mid{\buildrel{\rightarrow}\over{J}}\mid N >$ where ${\buildrel{\rightarrow}
\over {S_i}}$ is the spin of the $i^{th}$ nucleon.  ($\lambda$ can be expressed
in terms of the nuclear magnetic moment and nucleon g-factors.)  $M_{coh}$ is
proportional to $C_q$ and $M_{inc}$ is proportional to $A_q-B_q$, explicit
formulae being given in Ref. [5].
\medskip
\noindent
IV.  RESULTS
\smallskip
Eq. (11) allows one to divide dark matter detectors into two categories:  those
sensitive to the incoherrent (spin dependent) scattering due to a large value
of $\lambda^2$J(J+1), and those sensitive to the coherrent scattering.
Examples of ``incohererent detectors" are $^3He$ and $^{40}Ca$ $^{19}F_2$ with
CaF$_2$ the most sensitive detector.  Eqs. (11) show that $R_{coh}\sim M_N$ and
$R_{inc}\sim 1/M_N$ for large $M_N$, the additional $M_N^2$ factor in $R_{coh}$
arising from the $m_q$ factor in Eq. (6), i.e. roughly speaking, all the quarks
add coherrently to yield a $M_N$ factor in the amplitude.  The remaining
detectors considered here, $^{76}Ge+ ^{73}Ge,~ ^{79}Ga ^{75}As,~~ ^{23}Na
{}~^{127}I$ and $^{207}Pb$ are all of the ``coherrent" type with Pb being the
most sensitive since it is heaviest.

The dependence of the expected event rate on the supergravity GUT parameters is
fairly complicated as each parameter enters in several places and the
constraint Eq. (1) on $\Omega_{\tilde Z_{1}}h^2$ limits the parameter space.
One can, however, get a qualitative picture of the parameter dependence by
studying several characteristic examples . Fig. 1 shows that R decreases
rapidly with $m_{\tilde g}$ (mainly because the ${\tilde Z_{1}}$ becomes more
Bino-like).  It also shows that R is larger for larger tan $\beta$.  (See e.g.
the 1/cos $\beta$ factor in the denominator of the d-quark part of Eq. (9); the
1/sin $\beta$ factor for the u-quark part never gets exceptionally large since
tan $\beta > 1$ in the radiative breaking scenario).  Finally we note that
R[Pb] is 5-10 times larger than R[CaF$_2$] which is also a general feature.
The tan $\beta$ dependence is shown more explicitly in Fig. 2 for the NaI and
Ge detectors.  (The three examples were chosen so the $\Omega_{\tilde
Z_{1}}h^2$ is roughly the same at each tan $\beta$ along each graph).  The NaI
curve lies higher than the Ge one for each pair since $^{127}I$ is heavier than
$^{76}Ge$.

In general, the event rate drops with increasing $m_o$, as one would expect
since the squark mass increases with $m_o$, reducing the effect of the
s-channel squark pole.  (There are additional effects, however, as $m_o$ also
enters in the radiative breaking equations, effecting the size of $\mu$.)  Fig.
3 illustrates the general behavior for several of the detectors.  The coherrent
detectors, Pb, NaI, Ge, scale almost exactly by their atomic numbers.  (Fig. 3
also exhibits one of the few regions of parameter space where the $CaF_2$
detector lies above the Pb detector.)

Fig. 4 exhibits the maximum and minimum event rates for the Pb detector (the
most sensitive of the coherrent detectors) and the $CaF_2$ detector (the most
sensitive of the incoherrent detectors) as a function of $A_t$, as all other
parameters  are varied over the entire space.  One sees that generally a Pb
detector will be a factor of 5-10 times more sensitive than a $CaF_2$ detector.
Other coherrent detectors have event rates that scale with the Pb curve in
proportion to their atomic number while the $^3He$ has event rates a factor of
3 smaller than CaF.

The above analysis has been done with $\Omega_{\tilde Z_{1}}h^2$ obeying the
bounds of Eq. (1).  We discuss now the effect of varying these upper and lower
limits.  As mentioned in Sec. III, the event rate R rises with decreasing
$\Omega_{\tilde Z_{1}}h^2$ and this rise is rapid near $\Omega_{\tilde
Z_{1}}h^2 \simeq 0.1$.  Further, the maximum value of R occurs when $m_{\tilde
g}$ is near its minimum value.  However, by the scaling relations Eq. (2) this
can force $m_{\tilde W_{1}}< 45 GeV$, and hence such parameter points are
excluded by the LEP bounds.  This is what causes the sharp peaks in Fig. 4,
which occur when $m_{\tilde W_{1}}$ lies just above the LEP cut.  If, for
example, one increases the lower bounds on $\Omega_{\tilde Z_{1}}h^2$ to 0.15,
one finds that the maximum event rates follow the curves of Fig. 4 with the
peaks cut off.

The upper bound on $\Omega_{\tilde Z_{1}}h^2$ determines the minimum event
rates.  This is because the minimum rates occur when $m_{\tilde g}$ takes on
its largest value.  As $m_{\tilde g}$ increases, so does $m_{\tilde Z_{1}}$ by
Eq. (2).  The ${\tilde Z_{1}}$ annihilation cross section then drops and
$\Omega_{\tilde Z_{1}}h^2$ rises.  The upper bound of Eq. (1) on
$\Omega_{\tilde Z_{1}}h^2$ is found to occur when $(m_{\tilde g})_{Max}\cong$
750 GeV.  If one reduces the upper bound on $\Omega_{\tilde Z_{1}}h^2$ to 0.2
(which is consistent with the inflationary scenario which prefers $h\simeq
0.5$).  Then the maximum value of $m_{\tilde g}$ is reduced to \footnote*{Such
a low mass gluino could make it accessible to detection at the Tevatron.}
($m_{\tilde g})_{Max}\simeq$ 400 GeV.  This then increases the minimum event
rate curves of Fig. 4 by about a factor of 10.
\medskip
\noindent
V.  DETECTION POSSIBILITIES
\smallskip
The above discussion has analysed the expected event rates for a variety of
dark matter detectors over the range of parameters of supergravity GUT models.
These detectors are most sensistive to the region of parameter space where tan
$\beta$ is large and $m_o$ and $m_{\tilde g}$ are small.  Two types of
detectors were noted:  those with nuclei most sensitive to the spin dependent
incoherrent scattering of the ${\tilde Z_{1}}$ (e.g. $CaF_2$), and those most
sensitive to coherrent scattering (e.g. Pb).  In general, the best of the
coherrent scatters are more sensitive than the incoherrent scatterers by a
factor of 5-10.  The coherrent scatterer event rates scale approximately with
their atomic number.

Dark matter detectors currently being built plan to obtain a sensitivity to
signals with $R\r 0.1$ events/kg da.  Future detectors may be able to obtain a
sensitivity of $R\r 0.01$ events/kg da.  From Fig. 4 one sees that the present
detectors will be able to study a small fraction of the total event rate,
particularly the large tan $\beta$ region.  A more sizeable portion could
become available to the next generation of detectors.  However, it would appear
that there will be regions of parameter space inaccessible to these types of
dark matter detectors.  In spite of this, the experimental study of even only a
small part of the parameter space is of real importance, as such results,
combined with other experiments (e.g. the $b\rightarrow s + \gamma$ decay), can
together significantly limit the allowed parameter space of supergravity grand
unification models.  Thus recently it has been demonstrated that the
experimental limits on $b\rightarrow s + \gamma$ from CLEO do indeed affect
relic density analyses [18].  An analysis of the event rates with the inclusion
of the CLEO constraint will be given elsewhere [19].
\vfill\eject
\centerline{FIGURE CAPTIONS}
\smallskip
\item{Fig. 1}  Event rates R vs $m_{\tilde g}$ for Pb (solid curve and $CaF$
(dashed curve) detectors, with $A_t/m_o$ = 2.0, $m_o$ = 100 GeV and $\mu <$0.
The upper line of each curve is for tan $\beta$ = 20, the lower line for tan
$\beta$ = 6.  The gap between the short branches of the curves (at $m_{\tilde
g}$ = 225 GeV) and the main branches are regions where $\Omega_{\tilde
Z_{1}}h^2 < 0.1$.

\item{Fig. 2}  Event rates R vs tan $\beta$ for NaI and GeV detectors for
$m_{\tilde g}$ = 275 GeV, $\mu >0$.  The solid curve is for $A_t/m_o$ = 0.0,
$m_o$ = 200 GeV, the dashed curve for $A_t/m_o$ =0.5, $m_o$ = 300 GeV, the
dash-dot curve for $A_t/m_o$ = 1.0, $m_o$ = 200 GeV.  The upper curve of each
pair of curves is for NaI, the lower for Ge.

\item{Fig. 3}  Event rates R vs $m_o$ for $A_t/m_o$ = 0.5, tan $\beta$ = 8,
$m_{\tilde g}$= 300 GeV, $\mu >$ 0.  The dashed curve is for $CaF_2$, and solid
curves in descending order for Pb, NaI and Ge.

\item{Fig. 4}  Maximum and minimum event rates vs. $A_t/m_o$ for Pb (solid) and
$CaF$ (dashed) detectors, for $\mu < 0$.  All other parameters are varied over
the whole space to obtain the maximum and minimum event rates.

\vfill\eject

\centerline{REFERENCES}
\smallskip
\item {1.} M.W. Goodman and E. Witten, Phys. Rev. {\bf D31} (1985) 3059.

\item {2.}  K. Greist, Phys. Rev. Lett. {\bf 62} (1988) 666; Phys. Rev. {\bf
D38} (1988) 2357.

\item {3.}  R. Barbieri, M. Frigini and G.F. Giudice, Nucl. Phys. {\bf B313}
(1989) 725.

\item {4.}  M. Srednicki and R. Watkins, Phys. Lett. {\bf B225} (1989) 140.

\item {5.}  J. Ellis and R. Flores, Phys. Lett. {\bf B263} (1991) 259; Phys.
Lett. {\bf B300} (1993) 175; Nucl. Phys. {\bf B400} (1993) 25.

\item {6.}  M. Kamionkowski, Phys. Rev. {\bf D44} (1991) 3021.

\item {7.}  M. Drees and M. Nojiri, Phys. Rev. {\bf D48} (1993) 3483.

\item {8.}  J. Ellis, G. Ridolfi and F. Zwirner, Phys. Lett. {\bf B257} (1991)
83, {\bf B262} (1991) 477; H.E. Haber and R. Hempfling, Phys. Rev. Lett. {\bf
66} (1991) 1815.

\item {9.}  P. Gondolo and G. Gelmini, Nucl. Phys. {\bf 360} (1991) 145; K.
Greist and D. Seckel, Phys. Rev. {\bf D43} (1991) 3191.

\item {10.}  R. Arnowitt and P. Nath, Phys. Lett. {\bf B299} (1993) 58; (E)
{\bf B303} (1993) 403.

\item {11.}  P. Nath and R. Arnowitt, Phys. Rev. Lett. {\bf 70} (1993) 3696.

\item {12.}  B.W. Lee and S. Weinberg, Phys. Rev. Lett. {\bf 39} (1977) 165;
D.A. Dicus, E. Kolb and V. Teplitz, Phys. Rev. Lett. {\bf 39} (1977) 168; h.
Goldberg, Phys. Rev. Lett. (1983) 1419.

\item {13.}  J. Ellis, J.S. Hagelin, D.V. Nanopoulos, K. Olive and M.
Srednicki, Nucl. Phys. {\bf B238} (1984) 453.

\item {14.}  A.H. Chamseddine, R.Arnowitt and P. Nath, Phys. Rev. Lett. {\bf
49}, 970 (1982).  For reviews see P. Nath, R. Arnowitt and A.H. Chamseddine,
``Applied N=1 supergravity" (World Scientific, Singapore 1984); H.P. Nilles,
Phys. Rep. {\bf 110} (184) 1.

\item {15.}  P. Langacker, Proc. PASCOS90, Eds. P. Nath and S. Reucroft (World
Scientific, Singapore 1990); J. Ellis, S. Kelley and D.V. Nanopoulos, Phys.
Lett. {\bf B249} (1990) 441; {\bf B260} (1991) 131; U. Amaldi, W. De Boer and
H. Furstenau, Phys. Lett. {\bf B260} (1991) 447; F. Anselmo, L. Cifarelli, A.
Peterman and A. Zichichi, Nuov. Cim. {\bf 104A} (1991) 1817; {\bf 115A} (1992)
581.

\item {16.}  R. Arnowitt and P. Nath, Phys. Rev. Lett. {\bf 69} (1992) 725; P.
Nath and R. Arnowitt, Phys. Lett. {\bf B289} (1992) 368.

\item {17.}  J. Gunion, H. Haber, G. Kane and S. Dawson, ``The Higgs Hunter's
Guide" (Addison-Wesley, Redwood City, 1990).

\item {18.}  P. Nath and R. Arnowitt, CERN-TH-7214/94; NUB-TH.3093/94,
CTP-TAMU-32/94; F. Borzumati, M. Drees and M. Nojiri, KEK-TH-400, DESY 94-096,
MAD/PH/835.

\item {19.}  P. Nath and R. Arnowitt, CERN TH.7363/94, NUB-TH-3098/94,
CTP-TAMU-38/94.

\vfill\eject
\end